\documentclass[preprint,authoryear,12pt]{elsarticle}
\fontencoding{T1}
\usepackage{amssymb}
\usepackage{graphicx}
\usepackage{epstopdf}
\usepackage{slashbox}
\usepackage{picture}
\usepackage{natbib}
\usepackage{subfigure}

\journal{ASTIN Bulletin}
\begin{document}

\begin{frontmatter}
\title{Pricing Weather Derivatives for Extreme Events}

\author{Robert J. Erhardt, Richard L. Smith}

\address{Department of Statistics and Operations Research\\
University of North Carolina at Chapel Hill\\
Chapel Hill, North Carolina 27599\\
erhardt@email.unc.edu}

\begin{abstract}
We consider pricing weather derivatives for use as protection against weather extremes.  The method described utilizes results from spatial statistics and extreme value theory to first model extremes in the weather as a max-stable process, and then use these models to simulate payments for a general collection of weather derivatives.  These simulations capture the spatial dependence of payments.  Incorporating results from catastrophe ratemaking, we show how this method can be used to compute risk loads and premiums for weather derivatives which are renewal-additive.\\
\end{abstract}

\end{frontmatter}

\section{Introduction}
Weather derivatives are contingent contracts whose payments are determined by the difference between some underlying weather measurement and a pre-specified strike value.  They provide a useful risk management tool for any party facing weather risk.  They also provide investments which are often uncorrelated with more traditional financial instruments, allowing investors to diversify.  The first weather derivative was developed in 1996, and by 1999 derivatives and their options were being traded on the Chicago Mercantile Exchange \citep{kunreuther09}.

\citet{richards04} give a list of 5 elements common to all weather derivatives.  These include (a) an underlying weather index, (b) a well-defined time period, (c) the weather station used for reporting, (d) the payment attached to the index value, and (e) the strike value which first triggers payment.  The intention is for the buyer of the derivative to be compensated by the seller for amounts which roughly correspond to actual business losses.  Ideally, these losses are perfectly correlated with the payments of the weather derivative, though in practice this is rarely achieved.  Tailoring the contract to the specific needs of one buyer reduces its general appeal in a secondary market, and thus lowers the value of the contract.

Weather derivatives offer benefits to the buyer and seller not found in traditional insurance.  The buyer does not need to have an insurable interest, nor do they need to demonstrate an actual loss to receive payment.  The loss payment itself is generally proportional to the difference between the weather index and strike value.  Furthermore, weather derivatives offer the buyer the opportunity to sell (or even buy back) the contract in a secondary market such as the Chicago Mercantile Exchange, or over the counter.  There are no such markets for traditional insurance products.  Derivatives are also in general more lightly regulated, and payments are often considered taxable.

From the seller's point of view, weather derivatives offer several advantages over traditional insurance.  Weather derivatives avoid the higher administrative and loss adjustment expenses of insurance contracts.  They also eliminate concern for moral hazard, morale hazard, and fraud, as the event triggering the payment is easily verified and completely beyond the buyer's control.  When used to insure crops, weather derivatives help reduce the perceived information asymmetry associated with crop insurance, wherein farmers often have more information about their individual risk than insurers \citep{goodwin95}.

In this paper we consider the use of weather derivatives to provide protection against high impact, low probability business losses caused by extremes in the weather.  Though not technically insurance, we model losses and price derivatives using actuarial techniques originally developed to price insurance.  The weather derivatives we consider define some payment $L = L(M; s, t)$, where $M$ is the unknown weather random variable, and $s$ and $t$ are the pre-specified strike and limit values (occasionally we only write $L$ or $L(m)$ to refer to the loss to simplify notation).  Examples of three types of derivatives with payments based on high exceedances include

\begin{enumerate}
\item $L = \alpha$ if $\{M \ge s\}$ and 0 otherwise
\vspace{3mm}
\item $L = \beta \cdot (m-s)$ if $\{M \ge s\}$ and 0 otherwise
\vspace{3mm}
\item $L = \beta \cdot (m-s)$ when $\{s \le M \le t\}$ and $L = \beta \cdot(t - s)$ when $\{M \ge t\}$, and 0 otherwise,
\end{enumerate}
where $\alpha$ and $\beta$ are dollar values, and $m$ is the realization of random variable $M$.  The first provides a flat payment whenever the event $\{M \ge s\}$ occurs, the second provides a proportional payment based on the difference $(m-s)$, while the third limits the total payment.  Unlike most derivatives, weather derivatives do not have an underlying tradable asset, and thus many pricing approaches based on financial theory are inappropriate.  \citet{jewson05} provide an excellent reference and discussion of pricing techniques for weather derivatives.  The pricing approach we take is based on computing expected losses and expected loss variability.  The first step is to compute the expected payout $E(L) = \int L(m)g(m)\,dm$, where $g(m)$ is the density function of weather variable $M$.  For the three derivatives, shown, expected payments are

\begin{enumerate}
\item $E(L) = \int_{s}^{\infty} \alpha \cdot g(m) \,dm = \alpha \cdot P(M \ge s)$
\vspace{3mm}
\item $E(L) = \int_{s}^{\infty} \beta \cdot (m-s) g(m) \,dm$
\vspace{3mm}
\item $E(L) = \int_{s}^{t} \beta \cdot (m-s) g(m)\,dm + \beta \cdot (t-s) \cdot P(M \ge t)$.
\end{enumerate}

When viewed from the point of view of insurance, the quantities $E(L)$ are the \textit{pure premiums} of the contracts.  In the absence of any expenses, profit, risk loadings and time-value financial considerations, this is the price of the contract for the buyer.  Such contracts are fairly straightforward to price once one has an accurate estimate of the density function $g(m)$ in the region where $m \ge s$.  Further, one can compute all second moments as

\begin{enumerate}
\item $E(L^{2}) = \int_{s}^{\infty} \alpha^{2} \cdot g(m) \,dm = \alpha^{2} \cdot P(M \ge s)$
\vspace{3mm}
\item $E(L^{2}) = \int_{s}^{\infty} \{\beta \cdot (m-s)\}^{2} g(m) \,dm$
\vspace{3mm}
\item $E(L^{2}) = \int_{s}^{t} \{\beta \cdot (m-s)\}^{2} g(m)\,dm + \{\beta \cdot (t-s)\}^{2} \cdot P(M \ge t)$,
\end{enumerate}
and from these compute the variance $\mbox{var}(L) = E(L^{2}) - (E(L))^{2}$.  This information is often incorporated into the premium by adding a risk load $R(L)$, which is added to the pure premium as $P = E(L) + R(L)$.  Risk loads are meant to account for the additional risk taken on by writing derivatives with larger variability of losses.  Commonly, risk loads are a function of the variance (or standard deviation) of loss \citep{feldblum90}.  \citet{mango98} describes several common risk loadings such as $R(L) = \lambda \cdot \sqrt{\mbox{var}(L)}$, or $R(L) = \lambda \cdot \mbox{var}(L)$, where $\lambda$ is a dollar amount chosen to satisfy some risk tolerance criteria.

Next, consider a portfolio of $K$ weather derivatives, with aggregate payment $L = L_{1} + ... + L_{K}$.  The expected aggregate payment is simply the sum of the individual expected payments,
\[
E(L) = \sum_{k=1}^{K} E(L_{k}).
\]
However, when we allow for possible dependence among contracts, the variance of the aggregate payment is
\begin{equation}
\mbox{var}(L) = \sum_{k=1}^{K} \mbox{var}(L_{k}) + \sum_{k = 1}^{K-1} \sum_{k'=k+1}^{K} 2\cdot \mbox{cov}(L_{k}, L_{k'})\\
\label{eq:rho}
\end{equation}
The first issue when pricing weather derivatives (extreme or otherwise) is to properly address the positive correlation among contracts, with its resulting impact on aggregate loss variability as shown in equation~\ref{eq:rho}.  It is easy to envision positively correlated payments in a portfolio of weather derivatives, since weather variables are often positively correlated in space.  This correlation implies the variance of the aggregate payment exceeds the sum of individual payment variances, and thus risk loadings priced individually would be insufficient for the portfolio as a whole.

Next, consider the challenges when focusing on extreme weather events.  Examples of such events may include the maximum daily temperature exceeding some high threshold, the minimum daily temperature falling below some low threshold, or the minimum monthly rainfall falling below some low threshold. These events can often be written as $\{\max Y \ge s\}$ or $\{\min Y \le s\}$ where $Y$ is some weather-related random variable and $s$ is a pre-specified strike value.  In all cases we are defining a contract not based on ``typical" weather patterns of temperature or precipitation, but on extremes.  What is needed to accurately price these contracts is the distribution function of extreme events.  Furthermore, when one considers a collection of $K$ derivatives defined at different locations, one must carefully consider if the dependence of extremes is different from the dependence exhibited by non-extreme events.  Putting the two issues together, what is needed to price weather derivatives for extreme events is a model that (1) directly targets extremes, and (2) properly incorporates the spatial correlation of weather extremes.  One could further extend this to models which incorporate time dependence of extremes as well.  However, in this paper we focus on the spatial dependence of weather derivatives for extremes, as that is the largest omission of current methodology.

We begin with the problem of pricing a single weather derivative in the first section. This is handled through the Generalized Extreme Value distribution \citep{embrechts99}, which is the only permissible limit of the maxima of independent, identically distributed univariate random variables.  We demonstrate the approach by pricing a weather derivative based on extreme summer temperatures in Phoenix, Arizona.  In the third section, we introduce models used in spatial statistics and spatial extremes, which will serve as a foundation for the fourth and fifth sections.  There, we extend our weather derivative pricing model to multiple locations through the use of max-stable processes.  These processes capture dependence in spatial extremes.  Through large numbers of simulations, we can estimate all marginal variances, covariances, and other quantities of interest when pricing a portfolio of weather derivatives.  This information is ultimately incorporated into risk loads added to the pure premiums.  From the method presented, pure premiums and risk loads for a collection of $K$ spatially dependent weather derivatives can be obtained.  An application is shown in section 6.

\section{Pricing a Weather Derivative for an Extreme Event at a Single Location}
\subsection{The Generalized Extreme Value Distribution}
We begin by introducing the model used for modeling maxima (minima can always be rewritten as $\min(Y_{1}, ..., Y_{n}) = - \max (-Y_{1}, ..., -Y_{n})$, so there is no loss in generality when one considers only maxima).  Let $Y_{1}, ..., Y_{n}$ be independent and identically distributed univariate random variables with some distribution function $F$, and let $M_{n} = \max(Y_{1}, ..., Y_{n})$ be the maximum.  If $M_{n}$ converges to a non-degenerate distribution under re-normalization as
\[
P \left(\frac{M_{n} - b_{n}}{a_{n}} \leq m \right) = F^{n}(a_{n}m + b_{n}) \rightarrow G(m) \mbox{ as } n \rightarrow \infty
\]
for some sequences $a_{n}$ and $b_{n}$, then $G$ must be a member of the Generalized Extreme Value (GEV) family, with distribution function
\begin{equation}
G(m) = \exp \left\{-\left(1+ \xi \frac{m - \mu}{\sigma}\right)_{+}^{-1/\xi} \right\}.
\label{eq:Generalized Extreme Value}
\end{equation}
Here $a_{+} = \mbox{max}(a,0),$ and $\mu, \sigma, $ and $\xi$ are the location, scale, and shape parameters, respectively \citep{coles01}.  The sign of the shape parameter $\xi$ corresponds to the three classical extreme value distributions: $\xi>0$ is Fr\'{e}chet, $\xi<0$ is Weibull, and $\xi \rightarrow 0$ (in the limit) is Gumbel.  The Fr\'{e}chet case corresponds to a heavy tailed distribution, Gumbel is intermediate, and Weibull has a bounded upper limit.

In practice, it isn't necessary to worry about specifying sequences $b_{n}$ and $a_{n}$ due to the property of \textit{max-stability}: If $Y_{1}, ..., Y_{n}$ are independent and identically distributed from $G$, then $\max(Y_{1}, ..., Y_{n})$ also has the same distribution with only a change in location and scale, as $G^{n}(y) = G(A_{n}y + B_{n})$ for constants $A_{n}$ and $B_{n}$.  A distribution is a member of the Generalized Extreme Value family if and only if it is max-stable \citep{leadbetter83}.  A practical consequence of this property is that if one changes the block size (from monthly maxima to annual maxima, for instance) and fits a new model, new estimates for the three GEV parameters ($\mu, \sigma, \xi)$ are obtained, but the model is still GEV.

A special case of the Generalized Extreme Value family is the unit-Fr\'{e}chet, with distribution function $G(m) = \exp(-{1}/{m})$.  Any member of the Generalized Extreme Value family may be transformed to have unit-Fr\'{e}chet margins as follows: if $Y$ has a Generalized Extreme Value distribution with range $0<Y< \infty$, then a new variable $U$ may be defined as
\begin{equation}
U = \left(1 + \xi\frac{Y - \mu}{\sigma}\right)_{+}^{1/\xi}
\label{eq:U}
\end{equation}
and $U$ has unit-Fr\'{e}chet margins.  If the parameters are unknown, they may first be estimated and then the transformation to $U$ is taken.  When we model multivariate or spatial extremes, there is no loss in generality when one assumes the margins are all unit-Fr\'{e}chet.  In practice, one would first estimate all marginal distributions and transform to unit-Fr\'{e}chet, then in a second step analyze the spatial dependence.

The GEV model can be fit to observed data using maximum likelihood estimation.  Call the parameter vector $\phi$.  This parameter can be as simple as three fixed parameters, as $\phi = (\mu, \sigma, \xi)$.  Alternatively, one can model the GEV parameters using temporal or spatial covariates.  A few examples include $\mu = \mu_{1} + \mu_{2}\cdot t$, where $t$ is time, or $\sigma = \sigma_{1} + \sigma_{2}\cdot lat + \sigma_{3} \cdot lon + \sigma_{4} \cdot elev$, which considers effects of latitude, longitude, and elevation on the scale parameter.  No matter the structure of the parameter $\phi$, define the density function $g(m; \phi) = \frac{d}{dm}G(m; \phi)$.  Then, the maximum likelihood estimate of $\phi$ is
\begin{equation}
\hat{\phi}_{MLE} = \mbox{argmax}_{\phi} \prod_{i} g(m \mid \phi)
\label{eq:hatphi}
\end{equation}
This maximization is often done numerically, and has been implemented in a number of software programs including \texttt{R} \citep{R} using the function \texttt{fgev} in the package \texttt{evd}.  The density function for the fitted model is obtained by plugging in the maximum likelihood estimate as $g(m; \hat{\phi}_{MLE}).$

\subsection{Pricing a Contract Through Simulations}
Once we have a fitted model, we can use this to estimate the necessary pure premium and risk loading.  This requires estimates of the first two moments of the unknown payment variable $L$, as
\[
E(L^{d}) =  \int L(m; s, t)^{d} g(m)\,dm
\]
where $L(m; s, t)^{d}$ is the loss payment for realization $M=m$ raised to the $d^{th}$ power ($d=1$ or $2$), and $g(m)$ is the density function of the maxima.  The first type of weather derivative discussed has $L(m; s, t)^{d} = \alpha^{d}$ for $m \ge s$, which means the integral can be evaluated exactly as
\begin{equation}
\alpha^{d} \cdot P(M \ge s) = \alpha^{d} \cdot \left(1 - G(s; \hat{\phi})\right)
\label{eqone}
\end{equation}

The second and third types of derivatives involve more complicated integrals, so we use monte carlo techniques to estimate them.  This approach can be used to estimate moments for other types of derivatives with even more complicated payment structures, and is thus the most general approach.  Here we draw a large iid sample $M_{i} \sim G(m)$ for $i=1, ..., I$, and for each draw we compute the payment $L(m_{i})$.  Assuming that $E(|L(M)|^{2})$ is finite, then by the Strong Law of Large Numbers as $I \rightarrow \infty$, sample means converge to the first and second moments as \citep{robert07}
\begin{equation}
\frac{1}{I} \sum_{i=1}^{I} L(m_{i}) \rightarrow E(L(M)) = \int L(m)g(m)\,dm \hspace{5mm} \mbox{(almost surely)}
\label{eq:sa}
\end{equation}
and
\begin{equation}
\frac{1}{I} \sum_{i=1}^{I} L(m_{i})^{2} \rightarrow E(L(M)^{2}) = \int L(m)^{2} g(m)\,dm \hspace{5mm} \mbox{(almost surely)}.
\label{eq:sa2}
\end{equation}
Furthermore, if the fourth moment is finite, as $\int L(m)^{4}g(m)\,dm < \infty$, then by the Central Limit Theorem we know that the sample average in equation~\ref{eq:sa} is asymptotically normal with variance $\mbox{var}(L(M))/I$, and the sample average in equation~\ref{eq:sa2} is asymptotically normal with variance $\mbox{var}(L(M)^{2})/I$.
We estimate the expected payments under the second and third contracts by drawing $M_{1}, ..., M_{I} \sim G(m \mid \hat{\phi})$, and compute $L(m_{i})$ and $L(m_{i})^2$ for each of draw using $I=1,000,000$ total draws.  This total number was chosen to provide highly accurate estimates within a reasonable time, and simulations showed it was sufficiently high to eliminate concern for purely numerical monte carlo error.  Sample averages of each converge to the theoretical first and second moments, which are used in the pricing model.  One example of a risk-loaded premium based on marginal variance is
\begin{equation}
\hat{P} = \hat{E}(L) + \hat{R}(L) = \frac{1}{I} \sum_{i=1}^{I} L(m_{i}) + \lambda \cdot \left[\frac{1}{I} \sum_{i=1}^{I} L(m_{i})^{2} - \left(\frac{1}{I} \sum_{i=1}^{I} L(m_{i})\right)^{2} \right]
\label{eqtwo}
\end{equation}
for some dollar amount $\lambda$, chosen to satisfy some risk tolerance criteria.

\subsection{Example: Extreme temperature in Phoenix, Arizona}

As an example of how this model may be used, consider pricing a weather derivative with payments whenever the maximum daily summer temperature in the city of Phoenix, AZ exceeds some high threshold $s$.  On June 26, 1990, Phoenix airport was forced to close because the temperature exceeded 122 degrees Fahrenheit.  Aircraft operating manuals did not provide information for takeoff and landing procedures in temperatures above 120 degrees Fahrenheit.  The closure caused the predictable sort of economic disruption which accompanies airport closures.  We envision a weather derivative as a useful tool in this situation.

\begin{figure}
\begin{center}
\includegraphics[width=3in, angle=-90]{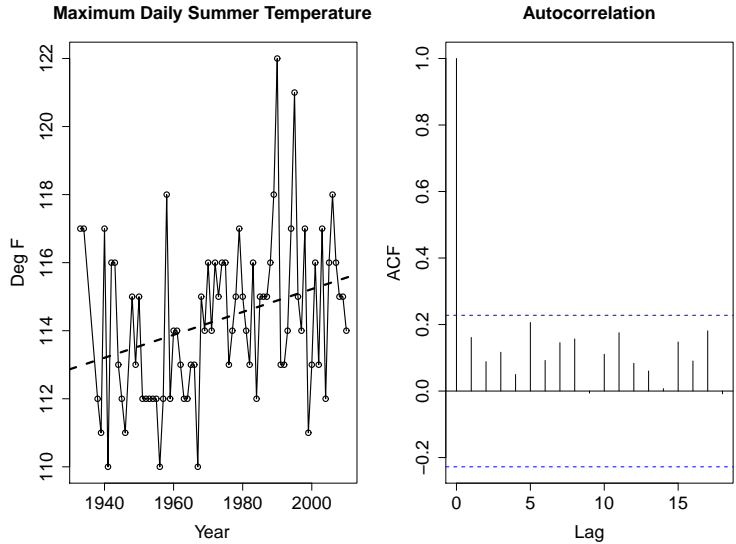}
\caption{Left: Maximum annual summer temperature at Phoenix International Airport from 1933-2010 (with some missing values).  The dashed line shows the annual trend, with statistically significant positive slope (p-val = 0.007).  Right: Empirical autocorrelation of maximum daily summer temperatures.  There is no evidence annual maximum daily temperatures are autocorrelated, as the value at all lags greater than 1 falls below the 95\% confidence interval line obtained from white noise sequences.}
\label{fig:phoenixtemp}
\end{center}
\end{figure}

To price the derivative, we collect maximum daily summer temperatures at the Phoenix airport, $y_{i, j}$ for year $i$ and day $j$, where $j=1, ..., 92$ (the 92 days in June, July, and August) for years 1933 to 2010.  This data comes from the National Climate Data Center.  For each year $i$, we take the block maximum $m_{i} = \max(y_{i,1}, ..., y_{i,92})$, and model these annual maxima $m_{i}$ as a Generalized Extreme Value distribution.  Plotting these data, we observe evidence of a slight positive trend over time (figure~\ref{fig:phoenixtemp}).  A simple linear model of maximum temperature versus year shows a statistically significant positive slope of 0.03363, with p-value 0.007.  We also find no evidence that annual maximum temperatures are autocorrelated.

\begin{figure}
\begin{center}
\includegraphics[width=4in, angle=-90]{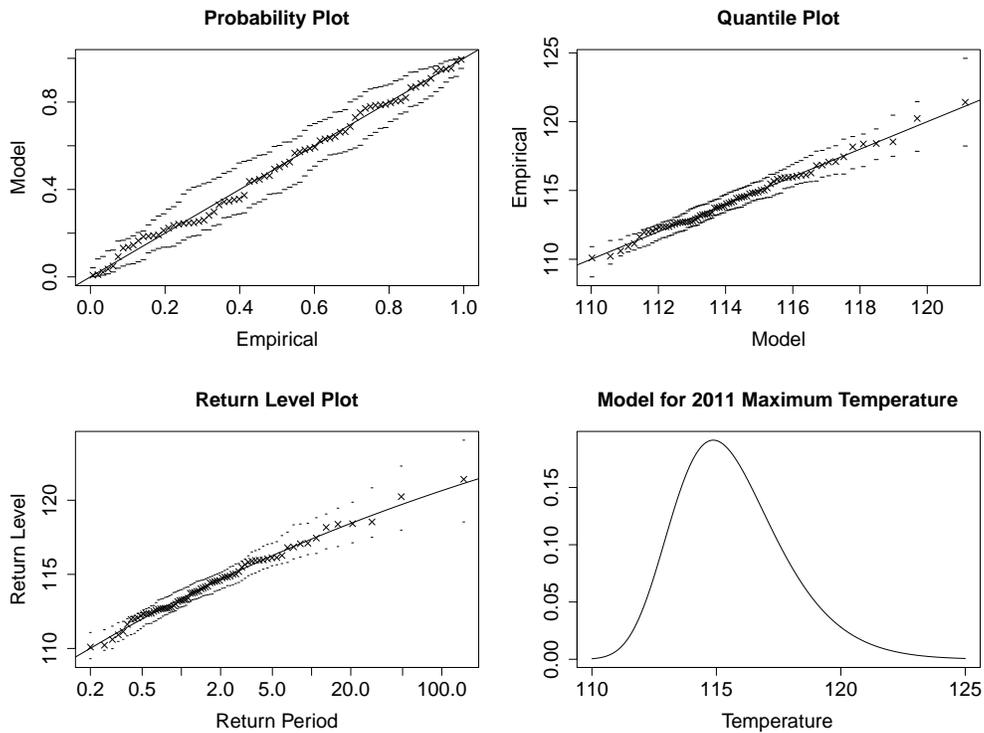}
\caption{Diagnostics from the maximum likelihood fit to the Phoenix summer temperature data.  Top left: comparison of empirical and model probabilities.  Top right: comparison of empirical and model quantiles.  Bottom left: The return period is the expected number of years required for the process to exceed the corresponding return level.  Bottom right: Model density function for 2011 maximum summer temperature in Phoenix AZ.}
\label{fig:phoenix}
\end{center}
\end{figure}

We estimate the GEV parameter $\phi$ using maximum likelihood estimation, as shown in equation~\ref{eq:hatphi}, but with the possibility of a trend on the location parameter, as $\mu = \mu_{1} + \mu_{2}\cdot t$ where $t$ is year.  Thus, the GEV parameter here is actually $\phi = (\mu_{1}, \mu_{2}, \sigma, \xi)$.  The maximum likelihood estimates (with standard errors shown in brackets) are $\hat{\mu}_{1} = 113.367 \hspace{2mm} [0.250]$, $\hat{\mu}_{2} = 0.035 \hspace{2mm} [0.011]$, $\hat{\sigma} = 1.931 \hspace{2mm} [0.176]$, and $\hat{\xi} = -0.090 \hspace{2mm} [0.078]$.  Figure~\ref{fig:phoenix} shows some common diagnostics and the return level plot.  The return level plot shows the expected number of years before an exceedance of a certain level is reached.  This is the same as the reciprocal of the probability of a specified exceedance, and forms the basis for statements such as describing an event as a``once every 50 years" event.

With the fitted model for maximum summer temperature, we can estimate the first and second moments of various weather derivative payments in the year 2011.  Estimated moments for the three types of derivatives from section 1 are shown in tables~\ref{table1},~\ref{table2},~\ref{table3}, and~\ref{table4}.  As the limit $t \rightarrow \infty$, payments under the second and third types are equal.

\begin{table}
\begin{center}
\caption{Phoenix, AZ example: Model first and second moments for the type 1 weather derivative with flat payment $L = 1000$ paid whenever the maximum daily temperature $M \ge s$ in the year 2011, using equation~\ref{eqone}.}
\begin{tabular}{c||c c c c c c}
Threshold $s$ & 114 & 116 & 118 & 120 & 122 & 124\\
\hline
$\hat{E}(L)$ & 759.11 & 391.84 & 144.11 & 41.61 & 9.72 & 1.79 \\
$\hat{E}(L^{2}) \cdot 10^{-3}$ & 759.11 & 391.84 & 144.11 &  41.61 &    9.72 &   1.79
\label{table1}
\end{tabular}
\end{center}
\end{table}

\begin{table}
\begin{center}
\caption{Phoenix, AZ example: Model first and second moments for the type 2 weather derivative with proportional payment $L = 1000 \cdot (m - s)$ in the year 2011, for varying thresholds of $s$.  Estimates are based on $I=1,000,000$ monte carlo draws used with using equations~\ref{eq:sa} and~\ref{eq:sa2}.}
\begin{tabular}{c||c c c c c c}
Threshold $s$ & 114 & 116 & 118 & 120 & 122 & 124\\
\hline
$\hat{E}(L)$ & 1,882.13 & 732.20 & 224.57 & 56.39 & 11.59 & 1.87\\
$\hat{E}(L^{2}) \cdot 10^{-3}$ & 7,336.56 & 2,369.34 &  627.82 &  137.98 &   24.45  &   3.26
\label{table2}
\end{tabular}
\end{center}
\end{table}

\begin{table}
\begin{center}
\caption{Phoenix, AZ example: Model first moments for the type 3 weather derivative with proportional payment $L = 1000 \cdot (m - s)$ up to limit $1000 \cdot(t - s)$ in the year 2011, for varying thresholds of $s$ and $t$.  Estimates are based on $I=1,000,000$ monte carlo draws used with using equations~\ref{eq:sa} and~\ref{eq:sa2}.}
\begin{tabular}{c||c c c c c c}
\backslashbox{t}{s} & 114 & 116 & 118 & 120 & 122 & 124\\
\hline
119 & 1,766.86 & 616.93 & 109.30 & & &\\
121 & 1,855.93 & 705.99 & 198.37 & 30.18 & & \\
123 & 1,877.34 & 727.41 & 219.78 & 51.59 & 6.80 & \\
125 & 1,881.44 & 731.51 & 223.89 & 55.70 & 10.90 & 1.18\\
$\infty$ & 1,882.13 & 732.20 & 224.57 & 56.39 & 11.59 & 1.87
\label{table3}
\end{tabular}
\end{center}
\end{table}

\begin{table}
\begin{center}
\caption{Phoenix, AZ example: Model second moments for the type 3 weather derivative with proportional payment $L = 1000 \cdot (m - s)$ up to limit $1000 \cdot(t - s)$ in the year 2011, for varying thresholds of $s$ and $t$.  Estimates are based on $I=1,000,000$ monte carlo draws used with using equations~\ref{eq:sa} and~\ref{eq:sa2}.  Values shown have order $10^{3}$.}
\begin{tabular}{c||c c c c c c}
\backslashbox{t}{s} & 114 & 116 & 118 & 120 & 122 & 124\\
\hline
119 & 5,894.14 & 1,383.33 &  98.21 &   &    &  \\
121 & 6,914.34 & 2,050.63 &  412.61 &  26.27 &   &  \\
123 & 7,243.24 &2,294.71 & 571.86 & 100.69 &   5.84  & \\
125 &  7,321.98 & 2,357.23 &  618.16 &  130.77 &   19.70 &    0.97 \\
$\infty$ & 7,336.56 & 2,369.34 & 627.82 & 137.98 &  24.45  &  3.26
\label{table4}
\end{tabular}
\end{center}
\end{table}

Tables like these can be used to price a wide range of weather derivatives.  Consider a weather derivative with payment $1000\cdot(M-118)$ for $M \le 125$ and $7000$ for $M \ge 125$, where $M$ is the maximum summer temperature in Phoenix.  Using equation~\ref{eqtwo} with $\lambda=0.0001$, the tables show the pure premium should be $223.89 + 0.0001 \cdot (618.16\cdot 10^{3} - 223.89^{2}) = 280.69$.  Premiums for other limits, strike values, and payment structures can be estimated from the same general approach once a fitted model has been obtained.

\section{Background on Spatial Statistics and Spatial Extremes}
To extend beyond a single location, we need to consider models for multivariate and spatial extremes.  Copulas provide a useful tool for modeling joint dependence, but they often fail to model extremes well \citep{mikosch06}.  Since weather has a natural spatial domain, a better choice is to build spatial models designed for extremes, and use those to determine the joint dependence in payments in a collection of weather derivatives.  In this section we present some background on spatial statistics, max-stable processes, and statistical methods for fitting max-stable processes to data.  The goal is to convey the benefits and motivation for using max-stable processes, and to outline the statistical method for fitting max-stable processes to data implemented in the \texttt{R} package \texttt{SpatialExtremes}.  Readers interested in more details of spatial statistics and max-stable processes can find much greater explanation in \citet{cressie93}, \citet{schlather02}, and \citet{padoan10}.

\subsection{Background on Spatial Statistics}
The basic object in spatial statistics is a stochastic process $Y(x), x \in X$ where $X$ is a subset of $R^{p}$, usually with $p=2$.  Let
\[
\delta(x) = E(Y(x)), \hspace{4mm} x \in X
\]
be the mean of the process defined for all of $X$, and assume that the variance of $Y(x)$ exists everywhere in $X$.  The process is said to be \textit{Gaussian} if for any $K \ge 1$ and locations $x_{1}, ..., x_{K}$, the vector $(Y(x_{1}), ..., Y(x_{K}))$ has a multivariate normal distribution.  The process is \textit{strictly stationary} if the joint distribution of $(Y(x_{1}), ..., Y(x_{K}))$ is the same as $(Y(x_{1}+h), ..., Y(x_{K}+h))$ for any $h \in X$ and for any $K$ points $x_{1}, ..., x_{K}$.  For a Gaussian process, strict stationarity implies
\[
\mbox{Cov}(Y(x_{1}), Y(x_{2})) = C(x_{1} - x_{2}) \hspace{3mm} \mbox{for all} \hspace{3mm} x_{1}, x_{2} \in X
\]
That is, the covariance of the process at any two locations is some function $C$ which depends only on the separation vector between points, and not the particular locations.  This is also called second-order stationarity.  Next, we define the \textit{variogram} through the relation
\[
\mbox{Var}(Y(x_{1}) - Y(x_{2})) = 2\gamma(x_{1}-x_{2})
\]
where the quantity $2\gamma$ is the variogram, and $\gamma$ is the \textit{semi-variogram}.  Under the assumption of strict (or second-order) stationarity,
\[
\gamma(h) = C(0) - C(h) = C(0)(1-\rho(h))
\]
where $\rho(h)$ is the correlation between two locations separated by vector $h$.
Further, if we have $\gamma(h) = \gamma(||h||)$ for all $h \in X$, meaning if the semi-variogram only depends on $h$ through its length $||h||$, then the process is \textit{isotropic}.  The correlation function $\rho(h)$ is then usually chosen from one of the valid families of correlations for Gaussian processes.  A few common choices are Whittle-Mat\'{e}rn, \[
\rho(h) = c_{1} \frac{2^{1-\nu}}{\Gamma(\nu)} \left(\frac{h}{c_{2}}\right)^{\nu}K_{\nu}\left(\frac{h}{c_{2}}\right), \hspace{3mm} 0 \leq c_{1} \leq 1, c_{2} > 0, \nu >0,
\]
Cauchy,
\[
\rho(h) = c_{1} \left\{1+\left(\frac{h}{c_{2}}\right)^{2} \right\}^{-\nu},  \hspace{3mm} 0 \leq c_{1} \leq 1, c_{2} > 0, \nu >0,
\]
and powered exponential
\[
\rho(h) = c_{1} \exp \left\{-\left(\frac{h}{c_{2}}\right)^{\nu}\right\} \hspace{3mm} 0 \leq c_{1} \leq 1, c_{2} > 0, 0 < \nu \leq 2,
\]
where $c_{1}, c_{2}$ and $\nu$ are the nugget, range, and smooth parameters, $\Gamma$ is the gamma function and $K_{\nu}$ is the modified Bessel function of the third kind with order $\nu$.  It is common to fix the nugget as $c_{1}=1$, which forces $\rho(h) \rightarrow c_{1} = 1$ as $h \rightarrow 0$.  This is a reasonable assumption for many environmental processes, and we make this assumption throughout this paper and do not attempt to model the nugget.  Throughout the remainder of this paper, the unknown spatial dependence parameter is called $\theta = (c_{2}, \nu)$.

\subsection{Multivariate Extreme Value Distribution}
The final definition we need before we can introduce the model for spatial extremes is the multivariate extreme value distribution.  Let $(Y_{i 1}, ..., Y_{i K})$, $i=1, ..., n$ be independent and identically distributed replicates of a $K-$dimensional random vector and let $M_{n} = (M_{n 1}, ..., M_{n K})$ be the vector of componentwise maxima, where $M_{n k} = \max(Y_{1 k}, ..., Y_{n k})$ for $k=1, ..., K$.  A non-degenerate limit for $M_{n}$ exists if there exist sequences $a_{nk} > 0$ and $b_{nk}$, $k=1, ..., K$ such that
\[
\lim_{n \rightarrow \infty} P \left(\frac{M_{n1}-b_{n1}}{a_{n1}} \leq m_{1}, ..., \frac{M_{nK}-b_{nK}}{a_{nK}} \leq m_{K} \right) = G(m_{1}, ..., m_{K}).
\]
Then $G$ is a multivariate extreme value distribution, and is max-stable in the  sense that for any $n \ge 1$ there exist sequences $A_{nk}>0$, $B_{nk}$, $k=1, ..., K$ such that
\[
G^{n}(m_{1}, ..., m_{K}) = G(A_{n1}m_{1} + B_{n1}, ...,A_{nK}m_{K} + B_{nK})
\]
The marginal distributions of a multivariate extreme value distribution are necessarily univariate Generalized Extreme Value distributions.

\subsection{Max-stable Processes}
Here we introduce the spatial analog of the multivariate extreme value distribution.  Let $Z(x), x \in X \subseteq R^{p}$ be a stochastic process.  If for all $n \ge 1$, there exist sequences $a_{n}(x), b_{n}(x)$ for some $x_{1}, ..., x_{K} \in X$ such that
\[
\lim_{n \rightarrow \infty} P \left(\frac{Z(x_{k}) - b_{n}(x_{k})}{a_{n}(x_{k})} \leq z(x_{k}), k = 1, ..., K \right) \rightarrow G_{x_{1}, ..., x_{K}}(z(x_{1}), ..., z(x_{K}))
\]
then $G_{x_{1}, ..., x_{K}}$ is a multivariate extreme value distribution.  If the above holds for all possible $x_{1}, ..., x_{K} \in X$ for any $K \ge 1$, then the process is a max-stable process.  To briefly summarize, if we have a max-stable process $Z(x)$ defined for all $x \in X$, then at any single location $x \in X$ the distribution of $Z(x)$ is GEV, and all finite vectors $(Z(x_{1}), ..., Z(x_{K}))$ follow a multivariate extreme value distribution.  In this sense, max-stable processes are the infinite dimensional generalization of multivariate extremes, and nicely extend the GEV to spatial domains.

Constructing a max-stable process is accomplished through a point process approach.  Let $Y(x)$ be a non-negative stationary process on $R^{p}$ such that $E(Y(x)) = 1$ at each $x$.  Let $\Pi$ be a Poisson process on $R_{+}$ with intensity $d w/w^{2}$.  If $Y_{i}(x)$ are independent replicates of $Y(x)$, then
\[
Z(x) = \max_{i} w_{i} Y_{i}(x), \hspace{5mm} x \in X
\]
is a stationary max-stable process with unit Fr\'{e}chet margins \citep{dehaan84}.  From this, the joint distribution may be represented as
\begin{equation}
P(Z(x) \leq z(x), x \in X) = \exp \left\{-E \left(\sup_{x \in X}  \frac{Y(x)}{z(x)} \right)\right\}
\label{eq:full}
\end{equation}

\begin{figure}[ht]
\centering
\includegraphics[width=5in]{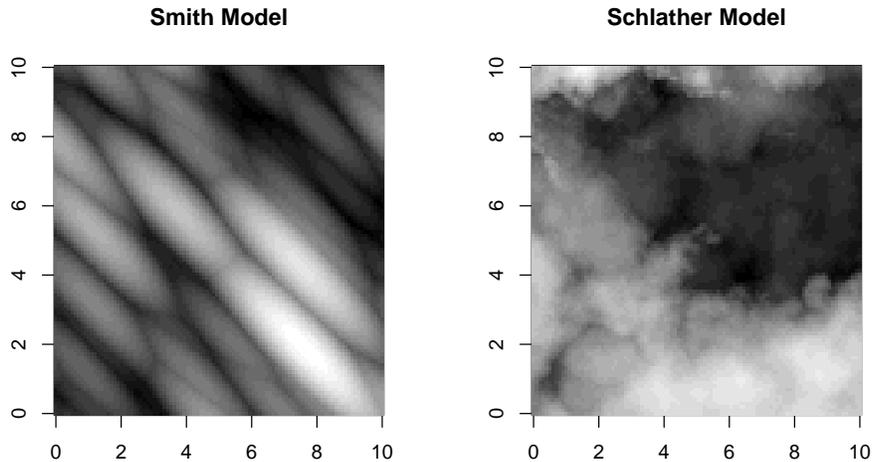}
\caption{Left: A realization of a Smith process with strong anisotropy.  Right: A realization of a Schlather process with Whittle-Mat\'{e}rn correlation and parameter $\theta = (c_{1}=1, c_{2}=3, \nu=1)$.}
\label{fig:maxstable}

\end{figure}

Varying the choice of the process $Y(x)$ gives different max-stable processes.  Smith (unpublished manuscript, 1990) constructed a process known as the Gaussian extreme value process by taking $Y_{i}(x)$ to be a multivariate Gaussian centered at the point $x_{i}$ with covariance matrix $\Sigma$.  Smith also introduced the ``rainfall-storms" interpretation in 2 dimensions: think of $R^{2}$ as the space of storm centers, $s_{i}$ as the magnitude of the $i^{th}$ storm, and $Y_{i}(x)$ as the shape of the storm centered at position $x_{i}$.  The maximum of independent storms at each location $x$ is taken to be the max-stable process.  A realization of this process is shown in figure ~\ref{fig:maxstable}.  A particular strength of the Smith model is the ability to handle anisotropy, as shown in the figure.  This comes from the off diagonal covariance parameter $\Sigma_{12}$ in
\[
\Sigma = \left[\begin{array}{cc}
\Sigma_{11} & \Sigma_{12} \\
\Sigma_{12} & \Sigma_{22}
\end{array}
\right]
\]
We will consider the use of the Smith model only to check the assumption of isotropy, which is required for the next class of max-stable processes. \citet{schlather02} introduced a flexible set of models for max-stable processes, termed extremal Gaussian processes.  Consider a stationary Gaussian process $Y(x)$ on $R^{p}$ with correlation function $\rho(\cdot; \theta)$ and finite mean $\delta = E \max(0, Y(x)) \in (0, \infty)$.  Let $w_{i}$ be a Poisson process on $(0, \infty)$ with intensity measure $\delta^{-1}w^{-2}ds$.  Then
\[
Z(x) = \max_{i} w_{i} \max(0, Y_{i}(x))
\]
is a stationary max-stable process with unit-Fr\'{e}chet margins.  The bivariate distribution function is
\begin{equation}
P(Z_{1} \leq z_{1}, Z_{2} \leq z_{2}) = \exp \left(- \frac{1}{2} \left[\frac{1}{z_{1}} + \frac{1}{z_{2}} \right] \left[1 + \left\{1 - 2(\rho(h; \theta)+1)\frac{z_{1}z_{2}}{(z_{1}+z_{2})^{2}}\right\}^{1/2} \right] \right)
\label{eq:sch}
\end{equation}
where $\rho(h; \theta)$ is the correlation of the underlying Gaussian process $Y(x)$ and $h=||x_{1}-x_{2}||$.  Figure \ref{fig:maxstable} shows one realization of a process with the Whittle-Mat\'{e}rn correlation function.  These processes are flexible, and produce plausible realizations of environmental processes.  We have elected to concentrate on the Schlather model in this paper.

\subsection{Maximum Composite Likelihood Estimation for Max-stable Processes}
A potential stumbling block to using max-stable processes is that the closed-form expression of the joint likelihood shown in equation~\ref{eq:full} can only be written out for dimension one or two.  This means only the univariate likelihood (which is GEV) and bivariate likelihood function (as shown in equation~\ref{eq:sch}) are available in closed form.  When one considers the joint distribution of a max-stable process at three or more locations, a closed-form expression for the joint likelihood is unavailable.  One way to proceed with a likelihood-based approach is to substitute the composite likelihood for the unavailable joint likelihood.  The composite log-likelihood is defined as

\begin{equation}
\ell_{C}(\theta; z) = \sum_{n=1}^{N} \sum_{i=1}^{I-1} \sum_{j=i+1}^{I} \log f(z_{i, n}, z_{j, n}; \theta)
\end{equation}
where each term $f(z_{i, n}, z_{j, n}; \theta)$ is a bivariate marginal density function based on locations $i$ and $j$.  The two inner sums sum over all unique pairs, while the outer sums over the $N$ i.i.d. replicates.  Similar to the full likelihood function, the parameter which maximizes a composite log likelihood can be found, and is termed a \textit{maximum composite likelihood estimate}, or MCLE.  The maximum composite likelihood estimator is consistent and asymptotically normal \citep{lindsay88} \citep{cox04} as
\begin{equation}
\hat{\theta}_{MCLE} \sim N(\theta, \tilde{I}) \hspace{3mm} \mbox{with} \hspace{3mm} \tilde{I} = H(\theta)J^{-1}(\theta)H(\theta).
\label{eq:sand}
\end{equation}
where $H(\theta) = E(-H_{\theta} \ell_{C}(\theta; Z))$ is the expected information matrix, $J(\theta) = V(D_{\theta}\ell_{C}(\theta; Z))$ is the covariance of the score, and $H_{\theta}$ is the Hessian matrix, $D_{\theta}$ is the gradient vector, and $V$ is the covariance matrix.  In the setting with the full likelihood and MLE, we have $H(\theta) = J(\theta)$, but for the composite likelihood setting these matrices are not equal.

\cite{padoan10} used composite likelihoods to model the joint spatial dependence of extremes, and implemented their work in the \texttt{R} package \texttt{SpatialExtremes}.  The maximum composite likelihood estimator $\hat{\theta}_{MCLE}$ is found using the numerical optimizer command \texttt{optim}.  Estimates of the variances are found by plugging in $\hat{\theta}_{MCLE}$ into expressions for $H$ and $J$:

\[
\hat{H}(\hat{\theta}_{MCLE}) = -\sum_{n=1}^{N} \sum_{i=1}^{I-1} \sum_{j=i+1}^{I} H_{\theta} \log f(z_{n, i}, z_{n, j}; \hat{\theta}_{MCLE})
\]
\[
\hat{J}(\hat{\theta}_{MCLE}) = -\sum_{n=1}^{N} \sum_{i=1}^{I-1} \sum_{j=i+1}^{I} D_{\theta} \log f(z_{n, i}, z_{n, j}; \hat{\theta}_{MCLE}) D_{\theta} \log f(z_{n, i}, z_{n, j}; \hat{\theta}_{MCLE})^{T}
\]

Model selection is based on minimizing the \textit{composite likelihood information criteria} (CLIC) \citep{varin05}, equal to
\begin{equation}
-2 \ell_{C}(\widehat{\theta}_{MCLE}; Z) - \mbox{tr}\left(\hat{J}(\hat{\theta}_{MCLE}) \hat{H}(\hat{\theta}_{MCLE})^{-1}\right)
\label{eq:CLIC}
\end{equation}
where the second term is the composite log-likelihood penalty term.

In our setting, this approach is used as follows: we begin with $Y$ independent and identically distributed realizations of an observed set of spatial extremes data, for locations $x_{1}, ..., x_{K}$.  Thus there are $Y$ replicates and $K$ locations.  For each location $x_{k}$, we transform the GEV data to unit-Fr\'{e}chet margins by first estimating all GEV parameters $\hat{\mu}(x_{k}), \hat{\sigma}(x_{k}),  \hat{\xi}(x_{k}), k=1, ..., K$, then using these in the transformation shown in equation~\ref{eq:U}.  Next we obtain the maximum composite likelihood estimate for the dependence parameter $\widehat{\theta}_{MCLE}=(\hat{c}_{2}, \hat{\nu})$ of the max-stable process using the composite likelihood approach outlined above.  The result is a fitted model for the extremes at the $K$ specific locations, with spatial GEV parameter $\hat{\phi} = (\hat{\mu}(x_{k}), \hat{\sigma}(x_{k}),  \hat{\xi}(x_{k}), k=1, ..., K)$ and spatial dependence parameter $\hat{\theta} = (\hat{c}_{2}, \hat{\nu})$.

\section{Simulating Losses from Extremes at Multiple Locations}
From the fitted model, we can simulate a collection of max-stable process $Z_{i}(x_{k}), i=1, ..., I$ for the same $k=1, ..., K$ locations as the observed data, and then transform back to the original scale of extremes data by transforming margins using the estimated GEV parameter $\hat{\phi}$.  From this we can compute the payments from weather derivatives at each location as $L(m_{i, k}; s, t)$.

To price a collection of $K$ weather derivatives, with jointly dependent losses, there are several quantities of interest.  First, the total variability of loss payments is
\begin{equation}
\mbox{var}\left(\sum_{k=1}^{K} L_{k}\right) =  \sum_{k=1}^{K} \mbox{var}(L_{k}) + \sum_{k = 1}^{K-1} \sum_{k'=k+1}^{K} 2\cdot \mbox{cov}(L_{k}, L_{k'}).
\label{eq:var}
\end{equation}
Now consider a portfolio of $K-1$ derivatives, with the seller deciding whether or not to write a $K^{th}$ derivative.  The additional derivative will increase the total portfolio variance by
\begin{equation}
MV_{K} = \mbox{var}\left(\sum_{k=1}^{K} L_{k}\right) - \mbox{var}\left(\sum_{k=1}^{K-1} L_{k}\right) = \mbox{var}(L_{K}) + \sum_{k=1}^{K-1} 2\cdot \mbox{cov}(L_{k}, L_{K}).
\label{eq:MV}
\end{equation}
The covariance for any two derivatives at locations $x_{k}$ and $x_{k'}$ is
\begin{equation}
\mbox{cov}(L_{k}, L_{k'}) = E(L_{k}\cdot L_{k'}) - E(L_{k})E(L_{k'})
\label{eq:cov}
\end{equation}

Each of these quantities may be estimated from a large collection of $I$ simulations.  The total variance of the portfolio is
\begin{equation}
\widehat{\mbox{var}}\left(\sum_{k=1}^{K} L_{k}\right) = \left[ \frac{1}{I}\sum_{i=1}^{I}\sum_{k=1}^{K} L(m_{i,k})^{2} - \left(\frac{1}{I}\sum_{i=1}^{I}\sum_{k=1}^{K} L(m_{i,k})\right)^{2}   \right]
\label{eq:varest},
\end{equation}
the marginal variance for adding a $K^{th}$ derivative is
\begin{equation}
\widehat{MV_{K}} = \widehat{\mbox{var}}\left(\sum_{k=1}^{K} L(m_{i,k})\right) - \widehat{\mbox{var}}\left(\sum_{k=1}^{K-1} L(m_{i,k})\right)
 = \widehat{\mbox{var}}(L_{K}) + \sum_{k=1}^{K-1} 2 \cdot \widehat{\mbox{cov}}(L_{k}, L_{K}),
\label{eq:r}
\end{equation}
and the covariance between any two derivatives is
\begin{equation}
\widehat{\mbox{cov}}(L_{k}, L_{k'}) = \left(\frac{1}{I}\sum_{i=1}^{I}L(m_{i,k})\cdot L(m_{i, k'})\right) - \left(\frac{1}{I}\sum_{i=1}^{I}L(m_{i,k})\right)\left(\frac{1}{I}\sum_{i=1}^{I}L(m_{i,k'})\right)
\label{eq:covest}
\end{equation}

\subsection{Simulated Example}
We evaluate the approach through simulations, first with a single detailed case and then larger numbers of simulations.  It will be convenient to define some new notation to keep simulation results clear.  Call the true full parameter $\theta = (\mu(x_{1}), \sigma(x_{1}), \xi(x_{1}), ..., \mu(x_{K}), \sigma(x_{K}), \xi(x_{K}), c_{2}, \nu)$, and the estimated parameter $\hat{\theta}$.  Call the true marginal variance $MV(\theta)$, and call an estimate of this based on a fitted model $\widehat{MV}(\hat{\theta})$.  We begin with a single detailed case.

We simulated a max-stable process with parameters chosen to mimic annual temperature maxima in North America.  The process had unit-Fr\'{e}chet margins and Whittle-Mat\'{e}rn covariance with dependence parameter $(c_{1}, c_{2}, \nu) = (1,3,1)$ for 75 years at 20 locations randomly placed on a 10 by 10 grid.  Call the vertical dimension latitude ($lat$) and the horizontal longitude ($lon$).  To make this data consistent with annual temperature maxima, at each location we transformed to the GEV scale by specifying parameters $\mu(x) = 110 - lat/2$, $\sigma(x) = 1.5+lat/5,$ and $\xi(x)=-0.1$.  The basic idea was to imagine higher latitude locations having overall lower extreme temperatures, but higher variability of extremes.  We used these transformations for each of the 20 locations to produce a max-stable process with GEV($\mu(x), \sigma(x), \xi(x)$) margins.  We fix this as the ``observed" data.

Next, we analyzed these data using composite likelihood estimation.  For each location $x_{k}$, we obtained a maximum likelihood estimate $\hat{\phi}(x_{k}) = \hat{\mu}(x_{k}), \hat{\sigma}(x_{k}), \hat{\xi}(x_{k}), k=1, ..., K$ using equation~\ref{eq:hatphi}, and then used these to transform each margin to unit-Fr\'{e}chet using equation~\ref{eq:U}.  We fit a max-stable process with Whittle-Mat\'{e}rn correlation with nugget parameter 1, and obtained maximum composite likelihood estimate $(\hat{c}_{2}, \hat{\nu})$.  Using this fitted model, we simulated a large number of processes and transformed them to the temperature scale using $\hat{\mu}(x_{k}), \hat{\sigma}(x_{k}), \hat{\xi}(x_{k})$ at each location $x_{k}$.

Thus we have described a means of simulating $i=1, ..., I$ extreme temperature events $m_{i, k}$ for locations $x_{1}, ..., x_{K}$ from our fitted model.  This information was used to estimate payments for weather derivatives by computing $L(m_{i, k}; s, t)$ .  Table~\ref{table5} shows results from one simulation.  We simulated 200,000 extreme temperature events at the same 20 locations, and used these to compute payments $L_{i, k}$ for $i=1, ..., 200,000$ and $k=1, ..., K$.  Shown are payments for a weather derivative paying 1 when $T \ge 112$, the aggregate payment $\sum_{k=1}^{19} L_{k}$, and the payments for a possible weather derivative $L_{20}$.  The marginal variance of adding the $20^{th}$ derivative was estimated as $\widehat{MV}_{20}(\hat{\theta}) = 22.788 - 21.216 = 1.572$, which is clearly much larger than 0.081, the estimated variance of the payments when dependence terms are ignored.  An additional 200,000 simulations from the true model with parameter $\theta$ shows the true marginal variance $MV_{20}(\theta)$ is 1.250.  This particular simulation showed an overestimation of the marginal variance of 25.7\%, which is clearly a substantial error, but not when compared to the error from ignoring spatial dependence.
\begin{table}
\begin{center}
\caption{Simulated payments for weather derivatives paying 1 when ${T \ge 112}$ using the fitted model with parameter $\hat{\theta}$.  This information shows the marginal variance for adding the $20^{th}$ policy is $\widehat{MV}_{20}(\widehat{\theta}) = 1.572$.  Using the true model, $MV_{20}(\theta)$ is 1.250.  The difference comes from parameter estimation error in $\widehat{\theta}$.}
\begin{tabular}{c||c c c c c c c}
Event & $L_{1}$ & $L_{2}$ & ... & $L_{19}$ &$\sum_{k=1}^{19} L_{k}$ & $L_{20} $ & $\sum_{k=1}^{20} L_{k}$\\
\hline
1 & 0 & 0 & ... & 0 & 0 & 0 & 0\\
2 & 1 & 0 & ... & 1 & 8 & 1 & 9\\
3 & 1 & 1 & ... & 0 & 4 & 0 & 4\\
... & ... & ... & ...& ... & ... & ... & ...\\
200,000 & 0 & 0 & ... & 0 & 1 & 0 & 1\\
\hline
\hline
Mean & 0.095	&0.130 & ..	&0.189&	2.907&	0.089&	2.996\\
Variance & 0.086 &	0.113 & ... &	0.153 &	21.216 &	0.081 &	22.788
\label{table5}
\end{tabular}
\end{center}
\end{table}

\subsection{Simulation Study of Performance}
\begin{figure}
\begin{center}
\includegraphics[width=3.5in, angle=-90]{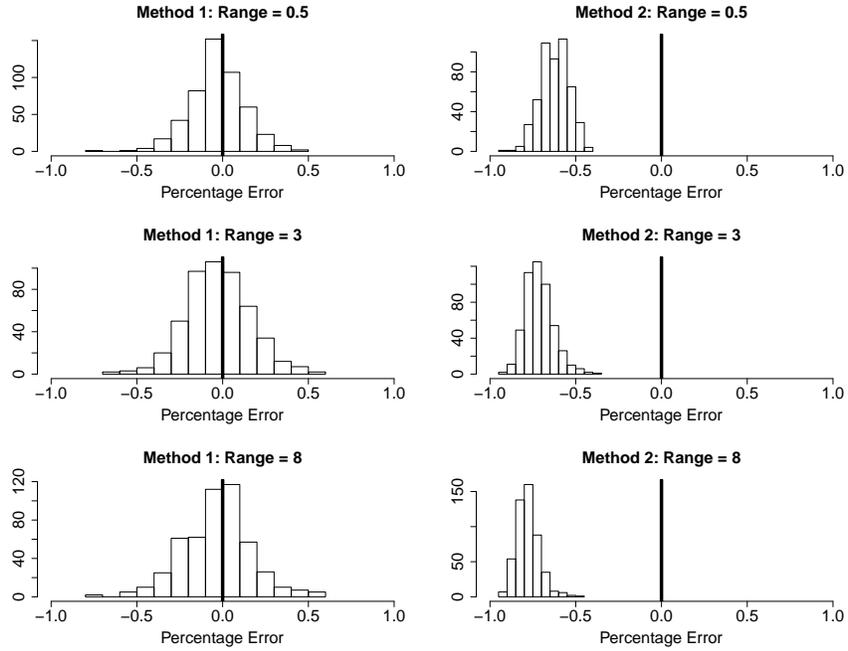}
\caption{Comparison of methods 1 and 2 for estimating the marginal variance $MV_{4}$ under three spatial dependence scenarios, with 500 simulations in each.  The heavy line at 0 signifies the true $MV_{4}(\theta)$.  The top row corresponds to a short-range dependence process, the middle row is medium-range, and the third row shows long-range spatial dependence.  The left column uses the approach outlined in this manuscript, also called Method 1, which incorporates spatial dependence.  The right column shows the non-spatial model in Method 2.  Results are plotted as percentage deviation from the true marginal variance, using equation~\ref{eq:mpe}.}.
\label{fig:compare}
\end{center}
\end{figure}

We evaluated the performance of this method in estimating the marginal variance of adding a $4^{th}$ weather derivative to an existing portfolio composed of $L_{1}, L_{2}$, and $L_{3}$.  This quantity is key to pricing a risk load for $L_{4}$.  We randomly placed $K=25$ locations on a 10 by 10 grid, and randomly selected 4 of these to represent locations of weather derivatives.  The target quantity was $MV_{4}$, the marginal variance of adding a fourth derivative.  We estimated this quantity using two methods:
\begin{enumerate}
\item Estimate $MV_{4}$ using equation~\ref{eq:r}, which accounts for spatial dependence by fitting a max-stable process and uses simulations from the model, with fitted parameter $\widehat{\theta} = (\hat{\mu}(x_{1}), \hat{\sigma}(x_{1}), \hat{\xi}(x_{1}), ..., \hat{\mu}(x_{4}), \hat{\sigma}(x_{4}), \hat{\xi}(x_{4}), \hat{c_{2}}, \hat{\nu})$.
\item Estimate $MV_{4}$ using equation~\ref{eq:varest}, which fits a GEV to the data at location $k=4$ but does not account for spatial dependence among the derivatives, with fitted parameter is $\widehat{\theta} = (\hat{\mu}_{4}, \hat{\sigma}_{4}, \hat{\xi})$.
\end{enumerate}
Again, call the true full parameter $\theta = (\mu(x_{1}), \sigma(x_{1}), \xi(x_{1}), ..., \mu(x_{K}), \sigma(x_{K}), \xi(x_{K}), c_{2}, \nu)$, and the estimated parameter $\hat{\theta}$.  Call the true marginal variance $MV(\theta)$, and an estimate $\widehat{MV}(\hat{\theta})$.  The true marginal variance was found by simulating $I=1,000,000$ realizations of a max-stable process under the true parameter $\theta$, and using equation~\ref{eq:r}.  Method 1 uses the same approach, but with estimated parameter $\hat{\theta}$, as we showed in the single example above.  Method 2 ignores spatial dependence.  The first measure of error we use is percentage error,

\begin{equation}
PE = \frac{\left(\widehat{MV}_{j}(\hat{\theta}) - MV_{j}(\theta)\right)}{MV_{j}(\theta)}
\label{eq:mpe}
\end{equation}
where $j=1, ..., 500$ refers to a simulation run.  This choice preserves the sign of estimation error.  Results are shown in figure~\ref{fig:compare}.  Here, we see the peril of ignoring spatial dependence of losses in a collection of weather derivatives.  The right column shows that as the range of the spatial dependence increases, the underestimation bias of estimating marginal variance $MV_{4}$ increases.  The left column shows the unbiased results obtained from incorporating dependence using the method of this paper.

We also show the asymptotic results for Method 1 in table~\ref{table6}.  Here, we use a slight variant of estimation error called mean absolute percentage error,

\begin{equation}
MAPE = \frac{1}{J} \sum_{j=1}^{J} \frac{|\widehat{MV}_{j}(\hat{\theta}) - MV_{j}(\theta)|}{MV_{j}(\theta)}.
\label{eq:mape}
\end{equation}

This choice does not preserve the sign of error, but is more suited to showing asymptotic results.  Results from 150 simulations in a variety of years and dependence ranges are shown in table~\ref{table6}.  For all dependence ranges shown, the error in estimation falls as more data is available.  The remaining error is primarily due to parameter estimation error in $\widehat{\theta}$ .

\begin{table}
\begin{center}
\caption{Mean absolute percentage error ($MAPE$) in estimating $MV_{4}$, the marginal variance of adding a fourth weather derivative to a portfolio (using only Method 1, which incorporates the spatial dependence).  Estimates are averages from 150 simulations each based on 25 locations, and are computed using equation~\ref{eq:mape} for 50, 100, 250, and 500 years of data.  For each spatial dependence range shown, error falls as the number of years of data increases.  The remaining error can be attributed to parameter risk.}
\begin{tabular}{c||c c c c}
\hline
Range $c_{2}$ & Y=50 & Y=100 & Y=250 & Y=500\\
\hline
Short 0.5 & 0.160 & 0.121 & 0.071 & 0.051\\
Medium 3 & 0.233 & 0.139 & 0.108 & 0.064\\
Long 8 & 0.231 & 0.151 & 0.087 & 0.064
\label{table6}
\end{tabular}
\end{center}
\end{table}

\section{Using Simulation Output to Price Weather Derivatives}

\subsection{Incorporating Marginal Variance into Risk Premiums}
Here we discuss using the models to price a risk load.  \citet{feldblum90} described five methods for an insurer to determine the risk load for writing a policy with unknown loss $L$.  The first two methods discussed were the variance approach, where $R = \lambda \cdot \mbox{var}(L)$, and the standard deviation approach, where $R = \lambda \cdot \sqrt{\mbox{var}(L)}$.  \citet{kreps90} and also \citet{philbrick91} discussed how a new policy adds risk through the \textit{marginal} increase in total variance, shown in equation~\ref{eq:r}.  \citet{gogol92} cautioned that these methods of pricing risk loads are \textit{order-dependent} if losses are correlated, leading to a mismatch between individual renewal risk loads and the total portfolio risk load, best illustrated by a toy example.

Consider two correlated policies, $L_{1}$ and $L_{2}$, and consider computing the risk load based on marginal variance in two ways:
  \begin{enumerate}
  \item A risk load using marginal variance for the total loss $L=L_{1}+L_{2}$ would be \\$\lambda \left(\mbox{var}(L_{1}) + \mbox{var}(L_{2}) + 2\cdot \mbox{cov}(L_{1}, L_{2}) \right)$.
  \item A risk load computed individually would go as follows: when $L_{2}$ is priced, the risk load is $\lambda(\mbox{var}(L_{2}) + 2\cdot \mbox{cov}(L_{1}, L_{2}))$.  When $L_{1}$ renews, it receives risk load $\lambda(\mbox{var}(L_{1}) + 2 \cdot \mbox{cov}(L_{1}, L_{2}))$.  The sum of these renewal risk loads is $\lambda \left(\mbox{var}(L_{1}) + \mbox{var}(L_{2}) + 4\cdot \mbox{cov}(L_{1}, L_{2}) \right)$, which has double counted the covariance terms and does not match the total portfolio risk load.
  \end{enumerate}
The example demonstrates the danger in careless accounting of covariance terms.  The approach we take to pricing a portfolio of dependent weather derivatives follows the work of \citet{mango98}.  In Mango's terminology, we use the covariance-share method, which apportions the total covariance between policies $L_{j}$ and $L_{K}$ and computes risk loads as
\begin{equation}
R(L_{K}) =\lambda \left(\mbox{var}(L_{K}) + 2 \sum_{j=1}^{K-1} a_{j, K} \cdot \mbox{cov}(L_{K}, L_{j}) \right)
\end{equation}
for any $0 \le a_{j, K} \le 1$.  This quantities $a_{j, K}$ are chosen to split the respective covariance terms and ensure the sum of individual renewal risk loads matches the total portfolio risk load.  One reasonable choice splits the total covariance in proportion to the expected losses of policies $j$ and $K$, as
\begin{equation}
a_{j, K} = \frac{E(L_{K})} {E(L_{j}) + E(L_{K})}.
\end{equation}
Under this choice, we always have $a_{j, K} + a_{K, j} = 1$, so the risk loads will always be renewal-additive.  Relevant quantities are estimated from large numbers of event simulations, and the risk load is
\begin{equation}
\hat{R}(L_{K}) =\lambda \left(\widehat{\mbox{var}}(L_{K}) + 2 \sum_{j=1}^{K-1} \hat{a}_{j, K} \cdot \widehat{\mbox{cov}}(L_{K}, L_{j}) \right)
\label{eq:rhat}
\end{equation}
using equations~\ref{eq:varest} and ~\ref{eq:covest}, where
\begin{equation}
\hat{a}_{j, K} = \frac{\frac{1}{I} \sum_{i=1}^{I} L(m_{i,K})} {\frac{1}{I} \sum_{i=1}^{I} L(m_{i,j}) + \frac{1}{I} \sum_{i=1}^{I} L(m_{i,K})}
\label{eq:a}
\end{equation}

\subsection{Example: Midwest Temperature Data}
\begin{figure}
\begin{center}
\includegraphics[width=2.5in,angle=-90]{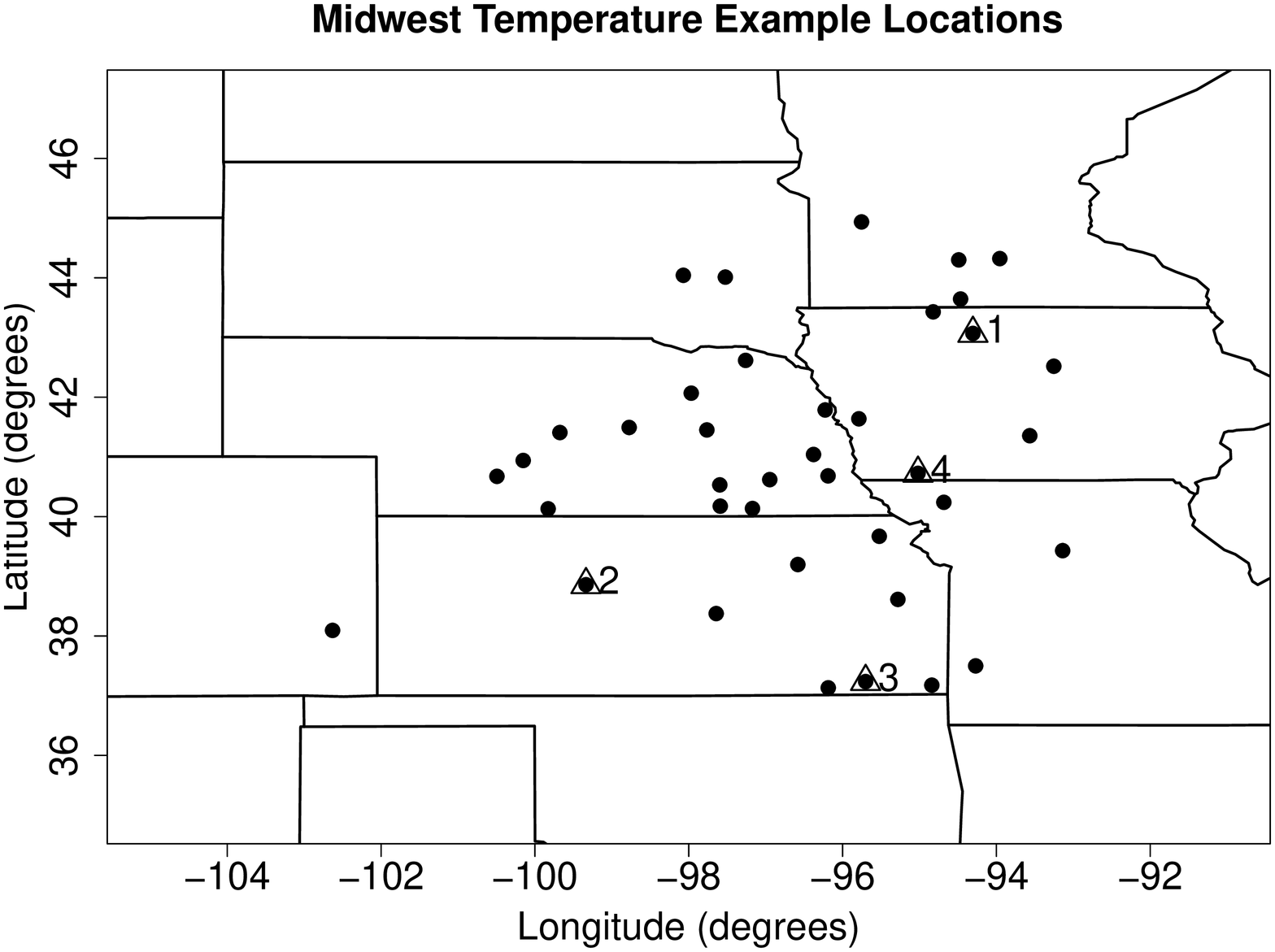}
\hspace{3mm}
\caption{Locations of the 39 stations in the Midwest temperature example used to fit the max-stable process to maximum summer temperature.  The locations in triangles labeled 1-4 are the places where weather derivatives are priced.}
\label{fig:region}
\end{center}
\end{figure}

We illustrate the methodology on US temperature data.  The data, freely available from the National Climate Data Center \verb (http://cdiac.ornl.gov/ftp/ushcn_daily/), come from 39 locations in the midwestern United States with complete summer (June 1 - August 31) temperature records from 1895 to 2009.  All sites are located between -93 and -103 degrees longitude, and 37 to 45 degrees latitude, shown in figure \ref{fig:region}.  We use all 39 locations to estimate the max-stable process, but we only consider weather derivatives at 4 of these locations, labeled 1-4 and drawn with triangles on the figure.  Call the maximum summer temperature at these $k=4$ locations $M_{i, k}$, with payments $L_{i, k}$ defined as

\begin{enumerate}
\item $L_{i, 1} = 1000$ if $\{M_{i, 1} \ge 107\}$ and 0 otherwise
\vspace{3mm}
\item $L_{i, 2} = 300 \cdot (M_{i, 2}-105)$ when $\{105 \le M_{2} \le 110\},$ $1500$ when $\{M_{i, 2} \ge 110\},$ and 0 otherwise
\vspace{3mm}
\item $L_{i, 3} = 200$ if $\{M_{i, 3} \ge 105\}$ and 0 otherwise
\vspace{3mm}
\item $L_{i, 4} = 200$ if $\{M_{i, 4} \ge 102\}$ and 0 otherwise
\end{enumerate}

The application proceeded in two steps.  The first is to use data from all 39 locations to fit a max-stable process in the study region, and the second is to then simulate temperature events from the fitted model only at locations 1-4 to estimate the renewal-additive risk load and premium for adding a weather derivative at location 4.

To investigate the possibility of a trend in maximum daily temperatures over time, we fit simple linear models to maximum daily temperature versus year, but found only 4 out of 39 locations showed statistically significant slopes at the $p=0.01$ level (this lower level was selected to reduce the false-positive rate which occurs with multiple tests).  Furthermore, all four slopes were negative.  We also fit GEV models to data from each station allowing for a time-varying location parameter as $\mu_{k} = \mu_{k,0} + \mu_{k,1}\cdot t$, where $t$ is year, but found only 7 differed significantly from 0 (again at the p=0.01 level), and again, all were negative.  These locations were spread throughout the study region, and showed no discernible spatial pattern or clustering.  We concluded that there was no evidence of a widespread shift in maximum temperatures over time throughout the entire region, and dropped the time-varying GEV location parameter.  However, just as a precaution we also conducted a separate analysis of the data including these 7 negative trends, but found it had little impact on the results.

We fit ordinary GEV models to each station, and obtained maximum likelihood estimate $\hat{\phi} = (\hat{\mu}(x_{k}), \hat{\sigma}(x_{k}), \hat{\xi}(x_{k}))$ for $k=1, ..., 39$.  Diagnostics like those shown in figure~\ref{fig:phoenix} gave no indication the GEV was inappropriate for any of these locations.  These fitted models were used to transform data at each location to unit-Fr\'{e}chet.  Next we assessed the appropriateness of using a max-stable process for the dependence.  We first fit a Smith process to the unit-Fr\'{e}chet data to check for anisotropy, but did not see strong evidence of anisotropy.  The parameter estimate of covariance $\widehat{\Sigma}$ were $\hat{\Sigma}_{11} = 2.064 \hspace{2mm} [0.020] \approx \hat{\Sigma}_{22} = 1.897 \hspace{2mm} [0.020]$, and $\hat{\Sigma}_{12} = -0.085 \hspace{2mm} [0.009] \approx 0$, where the number in brackets is the standard error of the estimate (when $\Sigma_{11} = \Sigma_{22}$ and $\Sigma_{12}=0$, we have perfect isotropy).  We next considered the more flexible Schlather model with Whittle-Mat\'{e}rn, Cauchy, and powered exponential correlation functions, and found the Whittle-Mat\'{e}rn to be the best with the lowest CLIC score.  Using the Whittle-Mat\'{e}rn correlation model, we obtained maximum composite likelihood estimates of the range and smooth as $\hat{c}_{2} = 4.6819 \hspace{2mm} [1.2975]$ and $\hat{\nu} = 0.3155 \hspace{2mm} [0.04625]$, where the number in brackets is the standard error of the estimate.

Next, we simulated $I=100,000$ max-stable processes from our fitted model at the four locations with weather derivatives.  Using the GEV estimates $(\hat{\mu}(x_{k}), \hat{\sigma}(x_{k}), \hat{\xi}(x_{k}))$ for $k=1, 2, 3, 4$ we transformed the unit-Fr\'{e}chet margins to GEV at each location.  Thus, we had simulations of maximum summer temperatures $M_{i,k}$ for $i=1, ..., 100,000$ at the $k=4$ locations.  From these, we computed the payments $L_{i, k}$ under the four contracts considered.  Table~\ref{table7} shows a few of these simulations.

\begin{table}
\begin{center}
\caption{Payments for weather derivatives in the Midwestern temperature example.  $I=100,000$ simulated extreme temperature events are simulated at locations 1-4 in figure~\ref{fig:region}.  The covariance share quantities $a_{j, K}$ are estimated using equation~\ref{eq:a}.}
\begin{tabular}{c||c c c c c c}
Event & $L_{1}$ & $L_{2}$ & $L_{3}$ &$\sum_{k=1}^{3} L_{k}$ & $L_{4} $ & $\sum_{k=1}^{4} L_{k}$\\
\hline
1 & 0 & 0 & 0 & 0 & 0 & 0\\
2 & 0 & 757.76 & 0 & 757.76 & 0 & 757.76\\
3 & 0 & 0 & 0 & 0 & 0 & 0\\
4 & 1000 & 964.02 & 0 &	1964.02 &	444.94 & 	2408.96\\
... & ... & ... & ...& ... & ... & ... \\
100,000 & 0 & 0 & 0 & 0 & 0 & 0 \\
\hline
\hline
Mean & 221.75 & 96.751 & 11.892 & 330.393 & 55.271 & 385.664\\
Variance ($\cdot 10^{-3}$)  & 172.58 &	99.89&	6.35&	381.38&	46.95& 561.98\\
$\mbox{Cov}(L_{k}, L_{4}) (\cdot 10^{-3}$) & 28.46 &	29.93 &	8.43 &	66.82 & &\\
$\hat{a}_{k, 4}$ & 0.1995&	0.3636	&0.8229& & &
\label{table7}
\end{tabular}
\end{center}
\end{table}

From equation~\ref{eq:rhat} and the information in the table, we compute the risk load for contract $L_{4}$ as
\[
\widehat{R}(L_{4}) = (46.95 + 2 \left(0.8229\cdot8.43 + 0.3636 \cdot 29.93 + 0.1995 \cdot 28.46 \right) )\cdot 10^{3} \cdot \lambda = 93,044 \cdot \lambda.
\]
This is roughly half of the total increase of $(561.98 - 381.38)\cdot10^{3}\cdot \lambda = 180,600 \cdot\lambda$.  The remainder would be apportioned to the risk loads of first three derivatives as they renew.

When we included the 7 time-varying location parameters, we computed a risk load of $90,914 \cdot \lambda$, a reduction of only 2.3\%.  This alleviates concerns that we might have wrongly ignoring trends in the GEV location parameter $\mu$.  If the inclusion of trends on the GEV location parameters $\mu_{k}, k=1, ..., 39$ had resulted in a substantially larger risk estimate, it might warrant the inclusion of trends as the more conservative choice.  However, the limited statistical evidence of trends combined with such a small reduction in the risk estimate supports dropping them altogether.

\section{Discussion}
We have described a means of pricing a collection of extreme weather derivatives based on simulations from max-stable processes.  Naturally, there will be some error between the collection of simulated payments and actual payments.  We discuss the errors introduced from model selection, simulations, and parameter estimation.

We have taken the approach to modeling spatial extremes as max-stable processes with Generalized Extreme Value margins, and naturally this model may not be appropriate for some spatial extremes data.  For weather derivatives with payments based on maxima (or minima) of some weather variable, models based on block maxima (or minima) of the data make the most sense, and certainly the GEV distribution has appealing asymptotic properties for these data.  \citet{coles01} discusses diagnostics to check the validity of the GEV for the marginal data.  Max-stable processes very naturally extend the GEV to the spatial domain, and are thus the logical choice for spatial block maxima data.  While a goal is to extend the approach presented in this manuscript to include non-stationary fields, at present this approach can only handle stationary fields.  Within the class of stationary max-stable processes, there are some choices of models.  One can model the GEV parameters $\mu, \sigma$, and $\xi$ with spatial, temporal, or other covariates, and one can consider different correlation functions $\rho(h)$ for the spatial dependence of the max-stable process.  In this paper we did not show much detail on model selection, however the paper by \citet{padoan10} shows the use of composite likelihood information criteria to handle model selection questions like these.

The computational cost of simulations from a max-stable process is minimal, and thus one can simulate hundreds of thousands or millions of events with relative ease.  Errors arising from numerical approximation in estimating the moments of payments assuming some fitted model using equations (6) and (7) are thus likely to be quite small, and can be made arbitrarily smaller with greater numbers of simulated events.

The largest source of error in this approach is likely to come from parameter risk - that is, the error in estimating the GEV and max-stable process parameters $\phi$ and $\theta$.  Reducing parameter risk is best handled through fitting the process to more weather data: more years of data, more locations of data, or ideally both.  A point worth stressing is that the data used to fit the max-stable process can (and probably should) contain far more locations than the portfolio of weather derivatives.  By adding additional points of data to fit the process, one reduces the parameter risk associated with estimating $\theta$, the spatial dependence parameter.

Our analysis is really a two-step procedure: the first transforms GEV margins to unit-Fr\'{e}chet by obtaining parameter estimate $\hat{\phi} = (\hat{\mu}(x_{1}), \hat{\sigma}(x_{1}), \hat{\xi}(x_{1}), ..., \hat{\mu}(x_{k}), \hat{\sigma}(x_{k}), \hat{\xi}(x_{k}))$, and then in a second step we fit a max-stable process to the transformed data to obtain dependence parameter estimates $\hat{\theta} = (\hat{c_{2}}, \hat{\nu})$.  We should point out that a single step procedure is possible, and is implemented in the package \texttt{SpatialExtremes}, but this has two drawbacks.  The first is that the numeric optimization of the likelihood needs to maximize a high dimension parameter.  In our example on Midwestern temperature data, the dimension would be $39 \cdot 3 + 2 = 119$ (and even larger if we kept time-varying GEV location parameters).  The dimension raises concerns that the numeric optimizer may converge to a local maxima, not the global one.  A second drawback is that single-step maximization of a max-stable process can be a painfully slow process, requiring orders of magnitude more time than a two-step procedure.  With these drawbacks in mind, the two-step procedure was selected.

One final comment is the potential mismatch between past and future weather extremes, particularly in the context of climate change.  One can model the location $\mu$ and scale $\sigma$ parameters of the GEV with time covariates to allow for the possibility of non-stationary maxima in time.  It is much less common to model the shape parameter $\xi$ as anything other than a fixed number.  We have illustrated the use of time covariates for modeling the location parameter as $\mu = \mu_{1} + \mu_{2} \cdot t$ in the Phoenix airport temperature example.  We caution readers not to extrapolate models such as these too far into the future.

\vspace{5mm}
Robert Erhardt\\
\textit{Department of Statistics and Operations Research}\\
\textit{University of North Carolina at Chapel Hill}\\
\textit{Chapel Hill, NC 27599}\\

\end{document}